\documentclass[a4paper, 11pt]{article}
\usepackage[cp1250]{inputenc}
\usepackage[margin=1.0in]{geometry}
\usepackage{rotating}
\usepackage{setspace,amsmath,amsfonts,amssymb}
\usepackage{graphicx}
\usepackage[usenames,dvipsnames]{xcolor}
\usepackage[round]{natbib}
\usepackage{xspace,helvet}
\usepackage{moreverb}        
\usepackage{url, booktabs}
\definecolor{Red}{rgb}{0.5,0,0}
\definecolor{NavyBlue}{rgb}{0.1,0.1,0.45}
\definecolor{MidnightBlue}{rgb}{0.1,0.1,0.65}

\usepackage[pdftex]{lscape}
\usepackage{fullpage}
\usepackage{booktabs}
\usepackage{multirow}
\usepackage{rotating}
\usepackage{pifont}
\usepackage{authblk}
\usepackage{setspace}
\usepackage{threeparttable}
\usepackage{tabularx}
\usepackage[toc,page]{appendix}
\usepackage{pbox}
\usepackage[font=small]{caption}

\newcommand{\rom}[1]{\uppercase\expandafter{\romannumeral #1\relax}}
\newcommand{\E}{\mathsf{E}}

\newcommand{\Prob}{\mathsf{P}}
\newcommand{\RNum}[1]{\uppercase\expandafter{\romannumeral #1\relax}}
\newcommand{\dee}{\,\mbox{d}}
\newcommand{\naive}{na\"{\i}ve }
\newcommand{\eg}{e.g.\xspace}
\newcommand{\ie}{i.e.\xspace}

\newcommand{\R}{\texttt{R}\xspace}

\begin{document}


\title{Event count distributions from renewal processes: fast computation
  of probabilities}
\author[1]{Rose Baker}
\author[2]{Tarak Kharrat}
\affil[1]{Salford Business School, University of Salford, UK.
rose.baker@cantab.net}
\affil[2]{Salford Business School, University of Salford, UK.
tarak.kharrat@gmail.com}
\date{\today}
\maketitle

\begin{abstract}
  Discrete distributions derived from renewal processes, \ie distributions of
  the number of events by some time $t$ are beginning to be used in econometrics
  and health sciences.
  A new fast method is presented for computation of the probabilities for
  these distributions.
  We calculate the count probabilities by repeatedly convolving the
  discretized distribution, and then correct them using Richardson extrapolation.
  When just one probability is required, a second algorithm is described,
  an adaptation of De Pril's method, in which the computation time does
  not depend on the ordinality, so that even high-order probabilities can be
  rapidly found. Any survival distribution can be used to model the inter-arrival times,
  which gives a rich class of models with great flexibility for modelling both
  underdispersed and overdispersed data.
  This work could pave the way for the routine use of these distributions
  as an additional tool for modelling event count data.
  An empirical example using fertility data illustrates the use of the method and
  was fully implemented using an \R \citep{Rlan} package \texttt{Countr}
  \citep{Countr} developed by the authors and available from the Comprehensive
  \R Archive Network (\texttt{CRAN}).
\end{abstract}
\section*{Keywords}
Renewal process; duration dependence; count data; Weibull distribution; convolution; 
Richardson extrapolation; hurdle model.

\section{Introduction}

Modelling a count variable (the number of events occurring in a given time interval) is 
a common task in econometrics. The standard approach is to use the Poisson model, where
$Y |x \sim \text{Poisson} (\E(Y|x) = \exp(x^\prime \gamma))$. Here $Y$ is predicted given
covariates with values $x$, using regression coefficients $\gamma$.
This model was built around
a one to one correspondence between the count model (Poisson) and the distribution
of the inter-arrival time (exponential). Perhaps this conceptual elegance contributed to
its popularity. With this elegance comes some limitation: the Poisson model
restricts the (conditional) variance to be equal to the (conditional) mean. This situation
is rarely observed in real life data and among the thousands of alternatives proposed in
the literature (see for example \citet{winkelmann2013econometric} or 
\citet{cameron2013regression} for a review), only a few retain the correspondence 
between the count model and the timing process. 

This correspondence is not only a conceptual
elegance but also offers the researcher the flexibility to model the aspect (counting or
timing) that he/she knows better (from the available data) and to draw
conclusions (typically prediction) using the other.
A very good example in the marketing context was given in \citet{shane}.

Another limitation of the Poisson model results from the memorylessness property of 
the exponential distribution. In fact, this property states that the probability
of having an arrival during the next $[t, t + \Delta t]$ time period
(where $t > 0$ and $\Delta t > 0$) is independent of when the last arrival occured.
In many situations, this assumption is not realistic and the history of the process
can be informative about future occurrences. For example, someone who consulted the 
doctor many times recently is more likely to have a higher number of doctor visits 
in the future (they are probably ill) than someone who did not. 
This is usually dealt with using the negative binomial model, where overdispersion is
accommodated by making the hazard of a series of visits of an individual a random variable
from a gamma distribution.

The distribution of $N(t)$, the 
number of renewal events by some time $t$ offers an alternative to the Poisson model that preserves the connection between the count
model and the timing process, but allows a more general event count distribution.  Inter-arrival times between events are still
assumed to be independent and identically distributed but the constant hazard function arising from an exponential distribution
is replaced by a nonconstant hazard function. These type of models display \textit{duration dependence} where
negative duration dependence is obtained by a decreasing hazard function (of time)
and positive duration dependence by an increasing hazard function. This gives a more flexible count distribution, and in particular,
allows it to be overdispersed or underdispersed.

It is possible to generalise further to a modified renewal process, which allows the time to the first event to have a different distribution from later event inter-arrival times.
This gives rise to a type of hurdle model (see \eg \citet{mullahy} for an account of hurdle models). 
If for example we kept the same survival distribution, but reduced the hazard function, we would have a distribution with an excess of zero events,
where the initial hazard function could be a different function of covariates from later ones. Conversely, if the initial hazard function is higher,
then we would see few zero events. Thus this class of distributions is flexible enough to analyse data with an abnormal number of zero events,
and often will have some foundation in reality.

In the simplest hurdle model, we have a Bernoulli trial, followed by a zero-truncated Poisson distribution for the number of events.
\citet[chapter 25]{greene} comments apropos of hurdle models that it is difficult to test whether the hurdle is really there or not (`regime splitting' is occurring), 
as the hurdle model cannot reduce to
the Poisson model and so give a nested model. However, modelling with a modified renewal process, we have to test only that the scale of the hazard function for the first event is
equal to that for the later events, when the hurdle model reduces to a regular model. This can be done with a chi-squared test derived from the log-likelihood function. Also, tests for under or overdispersion are difficult with hurdle models,
where the excess of zeros anyway induces overdispersion. With the modified Weibull process, a test for under or overdispersion even given a hurdle can be carried out
by using a chi-squared test based on the log-likelihood to test whether the shape parameter $\beta$ departs from unity. Renewal processes thus give rise to a rich and tractable class of models, but the slowness or unavailability of methods of computing the probabilities has so far largely prohibited their use.

\citet{winkl} was the first to comment on the usefulness of renewal process models
and derived a count model based on gamma distributed inter-arrival times. 
The choice of the gamma distribution was justified by computational necessity. In fact, the 
reproductive property of the gamma distribution, \ie sums of independent gamma
distributions are gamma distributed, leads to a simple form for the derived 
gamma count probability. 

The remainder of this paper is laid out as follows. We start by reviewing the
possible computation methods in Section~\ref{sec:possible}. Section~\ref{sec:allProba} 
discusses the situation when all probabilities up to the $m$th are required.
An alternative method is described in Section~\ref{sec:proba} when only the $m$th
probability is of interest, in which case a faster computation can be done.
Improvement by Richardson extrapolation is developed in Section~\ref{sec:Richar}.
Section~\ref{sec:general} contains a discussion on the generalisations 
to other survival distributions. In Section~\ref{sec:ex}, we re-analyse the 
same data used in \citet{winkl} and compare a sequence of nested models starting with
the basic Poisson regression. Using this approach allows us to highlight which features
of the model are most critical to describe the data at hand.
Future work and concluding remarks can be found in Section~\ref{sec:cl}.

\section{Possible computation methods for renewal processes}
\label{sec:possible}
In this section, we review the possible methods for computing the count
probabilities for other survival distributions besides the gamma.
\citet{lom} gave a method for computing a count model with Weibull interarrival times, based on an expansion of the exponential function into 
powers of $t$ and also into Poissonian functions.
\citet{shane} used the expansion into powers of $t$ to evaluate the discrete 
distribution probabilities and fit an underdispersed dataset (the one used
in \citet{winkl} and fitted here). The same approach has been used in \citet{jose2011count}
and \citet{Jose2013Gumbel} to derived a counting process with Mittag-Leffler 
and Gumbel inter-arrival times respectively. 

An expansion of the negative 
exponential is slow to converge. We found that this method can be 
improved by using techniques such as the Euler
and van-Wijngaarden transformations \citep[Chapter~5]{press}, which are designed to 
speed up convergence of alternating-sign series. Nevertheless, the convergence
is not guaranteed for probabilities of large numbers of events and
is not efficient if a high degree of accuracy is needed.

Throughout this paper we will use the Weibull distribution as our main example to
illustrate the methodology, which can be applied more generally. The survival function $P_0(t)$,
which is the probability of zero events by time $t$,
is given by $P_0(t)=\exp(-(\alpha t)^\beta)$.
This distribution allows both overdispersion ($\beta < 1)$ and 
underdispersion ($\beta > 1)$, and yields the Poisson distribution when $\beta=1$.
Before we develop our methodology to derive flexible count models based on renewal
processes, we first summarise the obvious available computational techniques that can be used. They are:

\begin{itemize}
\item expand out the exponential, using series transformations to speed up convergence. 
This is specific to the Weibull renewal process, but can be developed for others;
\item use (smart) Monte-Carlo simulation to generate renewal times up to time $t$ and 
read off the number of events $N(t)$;
\item use Laplace transforms, compute the survival distribution generating function, 
convert to the transform of the required probability, 
and invert the transform (\eg \citet{Chaudhry});
\item similarly, use the fast Fourier transform (FFT) which is often used for doing 
convolutions;
\item  evaluate the required probabilities directly as convolution integrals 
by discretizing the problem. This approach is the more attractive because \citet{depril} 
presented a recursive algorithm for computing the probabilities for the 
sum of $m$ discrete random variables, without computing the intermediate probabilities. 
\end{itemize}

The Monte-Carlo method is very easy to program, and useful for checking results of 
other methods. However, it cannot deliver high accuracy.
It can be made `smarter' by methods such as use of control variates,
antithetic variation, or importance sampling, but one really needs to
resort to Monte-Carlo simulation only for multidimensional integrals.
For univariate integrals evaluation by conventional quadrature methods is quicker 
and more accurate. For Weibull-like distributions,
the simple convolution method has error of $O(T^{-(1+\beta)/2})$,
where $T$ is computing time, whereas Monte-Carlo integration has error of $O(T^{-1/2})$, 
demonstrating that conventional quadrature is faster.
Note by the way that `error' in numerical integration is really what statisticians would 
call bias, rather than random error.

Convolution can be done directly, or via taking the Laplace or Fourier transform of 
the survival distribution pdf and inverting the result.
The drawback of directly doing convolutions
is that the time goes as $N^2$, where $N$ is the number of points into which
the probability is discretized. However, using Richardson extrapolation,
$N$ does not need to be very large, and so the advantage of transform methods largely disappears. 
The other advantage of transforms, that one can go straight to computation
of the $m$th probability, is removed by the availability of the
\citet{depril} method. It is perhaps also worth noting that
a quick look at transform methods throws up difficulties.
For example, the non-periodicity of the survival pdf gives an error
in the computed convolution. We have therefore used the direct method,
for which the size of errors is most easily considered; 
transform methods undoubtedly have potential but are not explored further here.

This paper focuses on the use of the discretized convolution method. 
To increase accuracy, Richardson extrapolation is used. The use of the trapezoidal rule,
together with Richardson extrapolation, is the basis of the well-known Romberg method 
of integration. Our approach is broadly similar.
The methodology described here could be applied (at least in outline) to any survival 
distribution, and hence is more general. The first part of our methodology, 
the discretized convolution, can indeed be applied to any distribution. 
The details of the second (extrapolation) step depend on the order of the error, 
and so will be specific to a distribution, or to a class of distributions.

\section{Computation of probabilities by convolution}
\label{sec:allProba}
Before discussing the convolution method and how it can be used to 
compute count probabilities, we recall the general framework used to build up the 
connection between the count model and inter-arrival timing process. Let 
$\tau_k, k \in \mathbf{N}$ be a sequence of \textit{waiting times} between the
$(k-1)$th and the $k$th event. The arrival time of the $m$th event is :
\[
  a_m = \sum_{\substack{k=1}}^m \tau_k, \ \ \ m=1,2,\dots 
\]
Denote by $N_t$ the total number of events in $[0, t)$. If $t$ is fixed, 
$N_t = N(t)$ is the count variable we wish to model. It follows that:
\[
N_t < m \Longleftrightarrow a_m \geq t
\]
Thus, if $F_m$ is the distribution function of $a_m$, we have
\[
\Prob(N_t < m) = \Prob(a_m \geq t) = 1 - F_m(t),
\]
Furthermore, 
\begin{align}
 \label{eq:rela} 
 \Prob(N_t = m) & =  \Prob(N_t < m + 1) - \Prob(N_t < m)  \nonumber \\
             & =  F_m(t) - F_{m + 1}(t) \\
             & =  P_m(t) \nonumber
\end{align}
Equation~(\ref{eq:rela}) is the fundamental relationship between the count variable
and the timing process. If the $\tau_k$  are iid with common density $f(\tau)$, the 
process is called a \textit{renewal process} (See \citet{feller1970} for a
formal definition). In this case, Equation~(\ref{eq:rela}) can be extended to obtain
the following recursive relationship:

\begin{align}
  P_{m+1}(t) & =  \int_0^t F_m(t-u)\dee F(u) - \int_0^t F_{m+1}(t-u)\dee F(u) \nonumber \\
            & =   \int_0^t P_m(t-u)\dee F(u),
\label{eq:conv}
\end{align}
where we have that $P_0(u)=S(u)=1-F(u)$, sometimes denoted the survival function.
Equation~(\ref{eq:conv}) can be understood intuitively: the probability of exactly $m+1$ 
events occurring by time $t$ is the probability that the first event occurs at time 
$0 \le u < t$, and that exactly $m$ events occur in the remaining time interval, 
integrated over all times $u$. Evaluating this integral, $P_1(t)\cdots P_m(t)$ can be 
generated in turn.

This is an attractive method of generating the required probabilities,
because the integrand is positive, so there are no subtractions
to increase rounding error.
To compute the integral, we use a method similar to the extended or composite
midpoint rule (\eg \citet[section 4.1.4]{press}). We have:
\[
\int_0^{Nh} f(x)\dee x = h\sum_{j=1}^N f\{(j-1/2)h\} + O(h^2),
\]
where there are $N$ steps with stepsize $h$, and $Nh=t$.
This is an open rule, \ie it does not require evaluating $f$ at the limits
of the integral. Thus 
\[
\int_{(j-1)h}^{jh} g(u)\dee F(u) =      \int_{(j-1)h}^{jh} g(u)f(u)\dee u 
                              \simeq g\{(j-1/2)h\}(F\{jh\}-F\{(j-1)h\}),
\]
where $g(u)=P_m(t-u)$ for some $m$, and $f$ is the pdf of the survival distribution. 
We make the choice of doing the integral of the pdf $f(u)$ analytically, so that
\begin{equation}
f((j-1/2)h)\simeq (F\{jh\}-F\{(j-1)h\})/h,
\label{eq:expandf}
\end{equation}
because this is simple for the Weibull distribution (and eventually other distributions) 
and increases accuracy to $O(h^{1+\beta})$.

The basic procedure is implemented in \texttt{getAllProbsUtil\_cpp()} function in the
\texttt{Countr} package \citep{Countr}. It generates probabilities $P_0\ldots P_m$.
On exit, the $P$ array (local) contains the probabilities $P_0 \cdots P_m$.
This code sets up $q$ (local) to contain $P_0$ at the midpoints $h/2\cdots (N-1/2)h$, 
sets up the $F\{jh\}-F\{(j-1)h\}$ array, and carries out the convolutions.
The array $q[~]$ initially contains $P_0$, and this is overwritten to contain $P_1$ etc. 

A crucial step is the shifting of the probabilities $q[k]$ left by $h/2$. 
This is necessary because $g$ must be used at the midpoint of each step, and the integral 
computes $g$ at the end of the step. With this correction, the result is $O(h^2)$
when $\beta \ge 1$, and $O(h^{\beta+1})$ for $\beta < 1$.
The algorithm uses $2N$ evaluations 
of the (Weibull) survival function (which is expensive) and then does
$(m-1)N(N+3)/2+N$ multiplications. Clearly, computing time increases as $N^2$ for large $N$.

\section{Computing one probability: adaptation of De Pril's method}
\label{sec:proba}
The method presented above computes all probabilities up to the $m$th,
which is slow if we need only the $m$th probability.
It can be improved so that computing time is $O(\ln(m)N^2)$ instead of $O(mN^2)$, using the addition chain method.
This is essentially an adaptation of a method that is used by compilers for fast computation of
integer powers of a variable with the minimum number of multiplications. The details are in Appendix~\ref{appendix:ch2A} .
This method, which we also call the `\naive method' is useful for timing comparisons, but our main interest is in the De Pril method, which can compute the $m$th probability in $O(N^2)$ operations.

\citet{depril} gave a method for computing the $m$-fold convolution of
a discrete distribution. He found that the idea dated back a long way,
being first used in other applications than probability before 1956.
We refer the reader to De Pril's paper for two derivations of this amazing
algorithm and its history, and simply present it here:
let $q_i$ be the value of probability density function of the survival distribution evaluated
 at points $t_i \ge 0$  where $q_0 > 0$. Then the probability of $m$ events
is $f_N^{(m)}$, the $m$-fold convolution of $q$,  given by 
\[f_0^{(m)}=q_0^m,\]
and for $N > 0$  by the recursion 

\begin{equation}
  f_N^{(m)}=q_0^{-1}\sum_{j=1}^N (\frac{(m+1)j}{N}-1)f_{N-j}^{(m)}q_j.
  \label{eq:depril}
\end{equation}
This algorithm when applied to our case requires three arrays:
one to hold the survival function, one for the probability mass $q$,
and one work array to hold $f$.

To apply this method to continuous distributions like the Weibull,
we first discretized the distribution, so that $q_j=F((j+1)h)-F(jh)$.
The probability mass $f_0^{(m)}$ has contributions from the $m$ random variables
all taking the value zero, up to them all taking the value $h-\epsilon$.
We should therefore estimate the mean as $mh/2$ rather than zero,
so an approximation to the continuous case is that all probability masses
such as the $N$th should be taken as pertaining to time $(N+m/2)h$.
To apply this continuity correction, we do not need to copy the $f_N^{(m)}$
into different array locations, but simply to reduce the time interval in
the survival function in \ref{eq:step2}). Finally, for even $m$, the latest
probability mass occurs exactly at time $t$, and so we take only half of this
probability mass. With these two crucial modifications, the method yields the same
accuracy as the earlier methods, and Richardson extrapolation can be applied as before.
The results are very similar to the addition-chain method, but are usually slightly more
accurate, and computation is of course faster.
An unexpected additional gain is that for even $m$, the survival function is not
required at half-integer values of $h$, so saving time on these computations.
It had been feared that the presence of the minus sign in the recursion
(\ref{eq:depril}) would degrade accuracy, but running the program in
quadruple precision gave identical results, so that is not a problem. 

Sometimes data are censored, and we only know that at least $m$ events have occurred. 
This probability $P_{\ge m}$ is then needed for likelihood-based inference.
For the direct method (Section~\ref{sec:allProba}), one would compute
$P_{\ge m}=1-\sum_{i=0}^{m-1}P_i(t)$, but for this method, which delivers $f_m(u)$, 
we compute $P_{\ge m}=\int_0^t f_m(u)\dee u$; the routine supplied in the \R 
package \texttt{Countr} returns this. 
This is an advantage of this and the addition chain method, because small probabilities obtained by 
differencing are subject to large errors.

The next section describes how Richardson extrapolation can be used to improve 
the accuracy, without necessitating a large value of $N$ and consequent slow computation.

\section{Improvement by Richardson extrapolation}
\label{sec:Richar}
In Romberg integration, the trapezoidal rule is used to generate approximations 
of error $O(h^2)$, and Richardson extrapolation is used to progressively
remove errors of order $h^2$, $h^4$ etc. Clearly, if an estimate 
$S_1=S+\gamma h^\delta$ and $S_2=S+\gamma (h/2)^\delta$, where $S_1$ and $S_2$ are the 
approximations with $N$ and $2N$ steps respectively and $S$ is the true value, 
we can remove the error and estimate $S$ as 
\begin{equation}
S_3=(2^\delta S_2-S_1)/(2^\delta-1)
\label{eq:rich}.
\end{equation}
Subsequently, higher-order errors can be removed in the same way until the 
required accuracy is attained. Romberg integration can also be done with the 
extended-midpoint rule (\eg \citet{press}).

\begin{figure}[b]
\centering
\makebox{\includegraphics{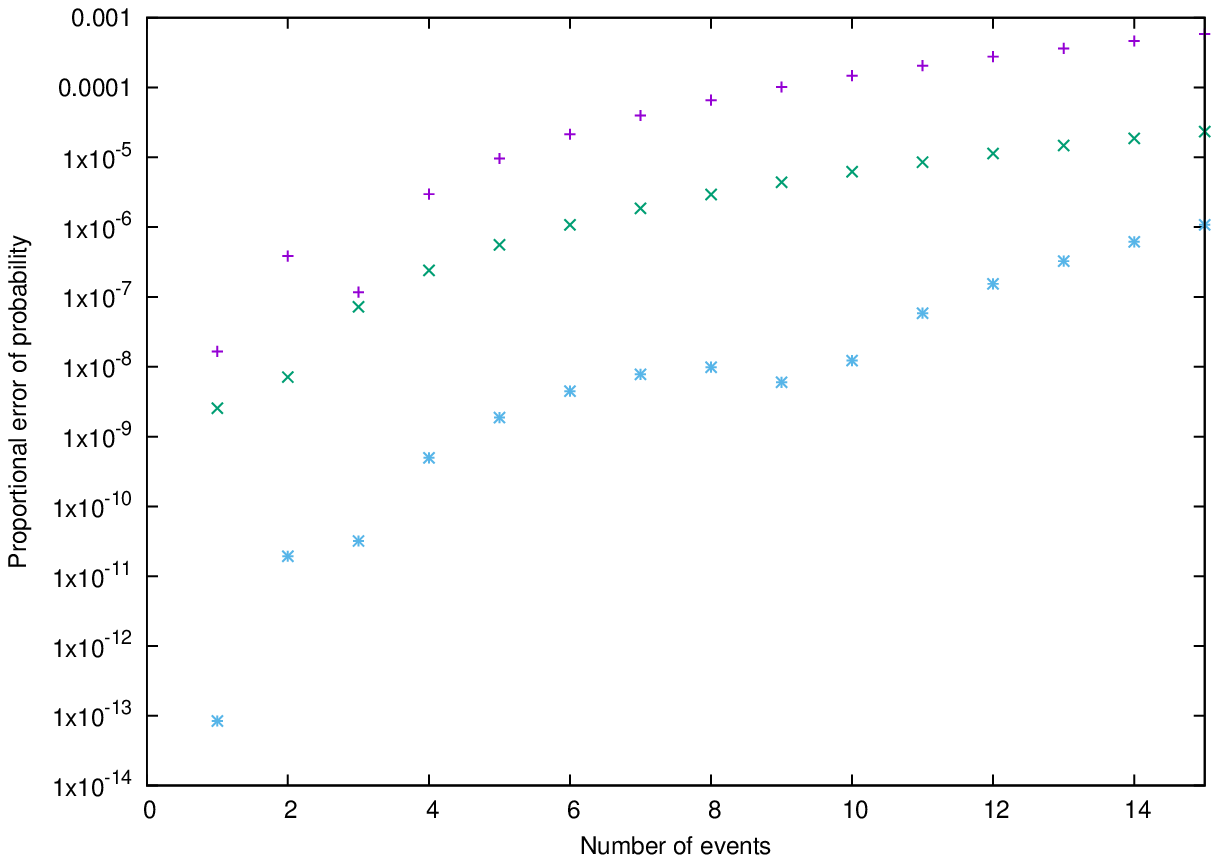}}
\caption{\label{figa} Proportional errors in probabilities for the \naive computation and the two Richardson corrections.
Here $\alpha=1, t=1, \beta=1.1$.}
\end{figure}

\begin{figure}
\centering
\makebox{\includegraphics{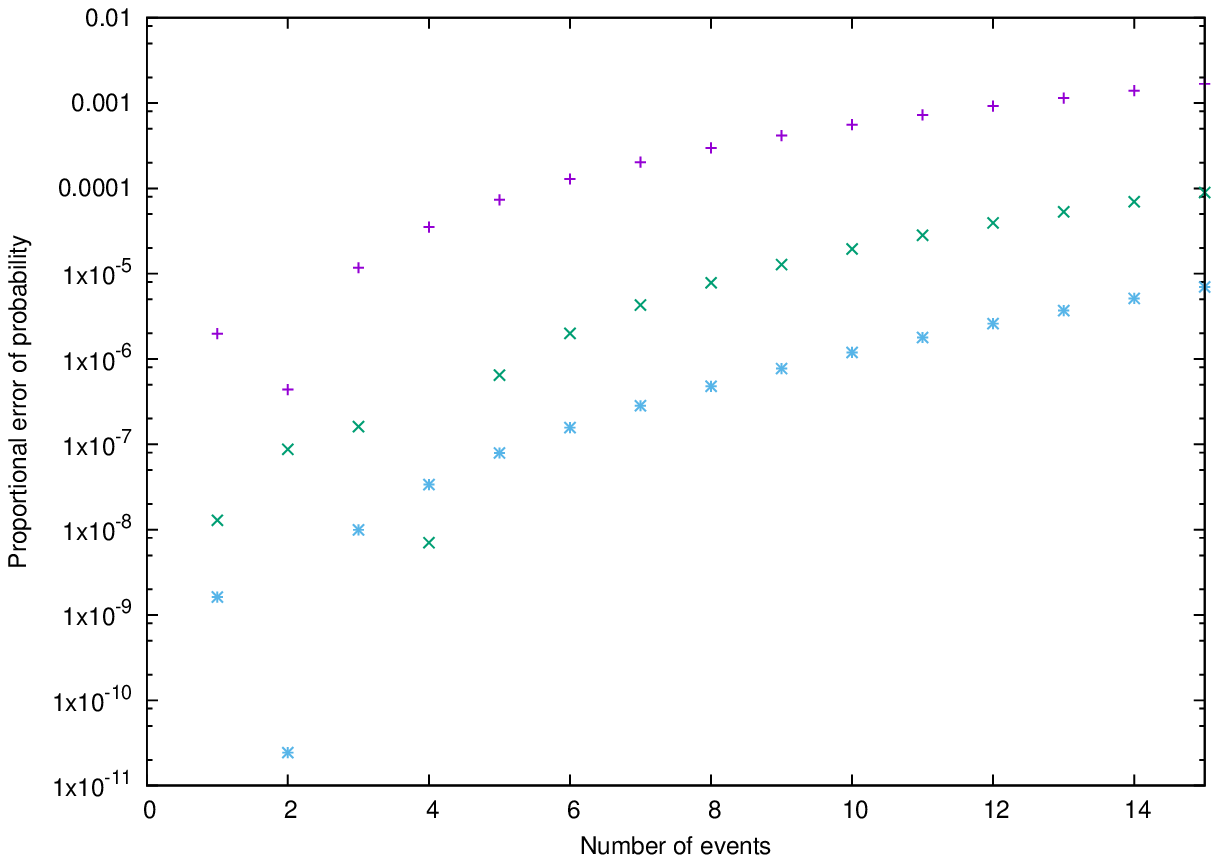}}
\caption{\label{figb} Proportional errors in probabilities for the \naive computation and the two Richardson corrections.
Here $\alpha=1, t=1, \beta=0.6$.}
\end{figure}

The situation for convolutions is less straightforward, but a satisfactory solution can 
be found, and the details are given in Appendix~\ref{appendix:ch2B}. 
We now study the proportional errors of probabilities, because these are
what determine the error in the in the log-likelihood.
Figure \ref{figa} shows absolute proportional errors $\delta p/p$ for the first 15
probabilities with $\beta=1.1$, for the \naive computation, after applying
a Richardson extrapolation for error $h^{1+\beta}$,and after applying
the second transformation to remove error $O(h^2)$.
It can be seen that the errors reduce substantially. Figure \ref{figb} shows
the estimated power of $h$ of the error, derived by applying (\ref{eq:gamma}),
with $\beta=1.2$. It can be seen that this is initially around 2
(because $1+\beta > 2$), and increases to 2.2, then to 3-4 after the second extrapolation.

\begin{figure}
\centering
\makebox{\includegraphics{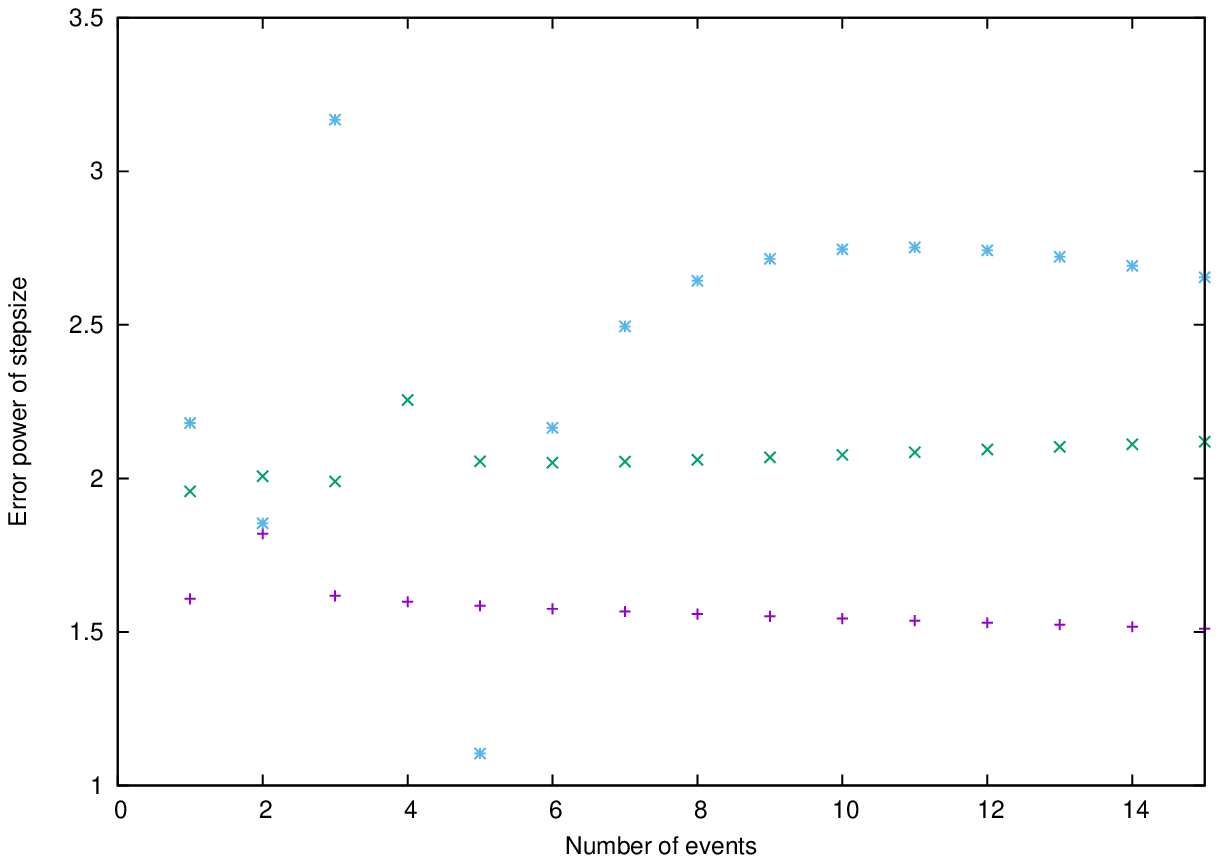}}
\caption{\label{fige} Powers of stepsize $h$ for error  in probabilities for the \naive computation and the two Richardson corrections.
Here $\alpha=1, t=1, \beta=0.6$.}
\end{figure}

\begin{figure}
\centering
\makebox{\includegraphics{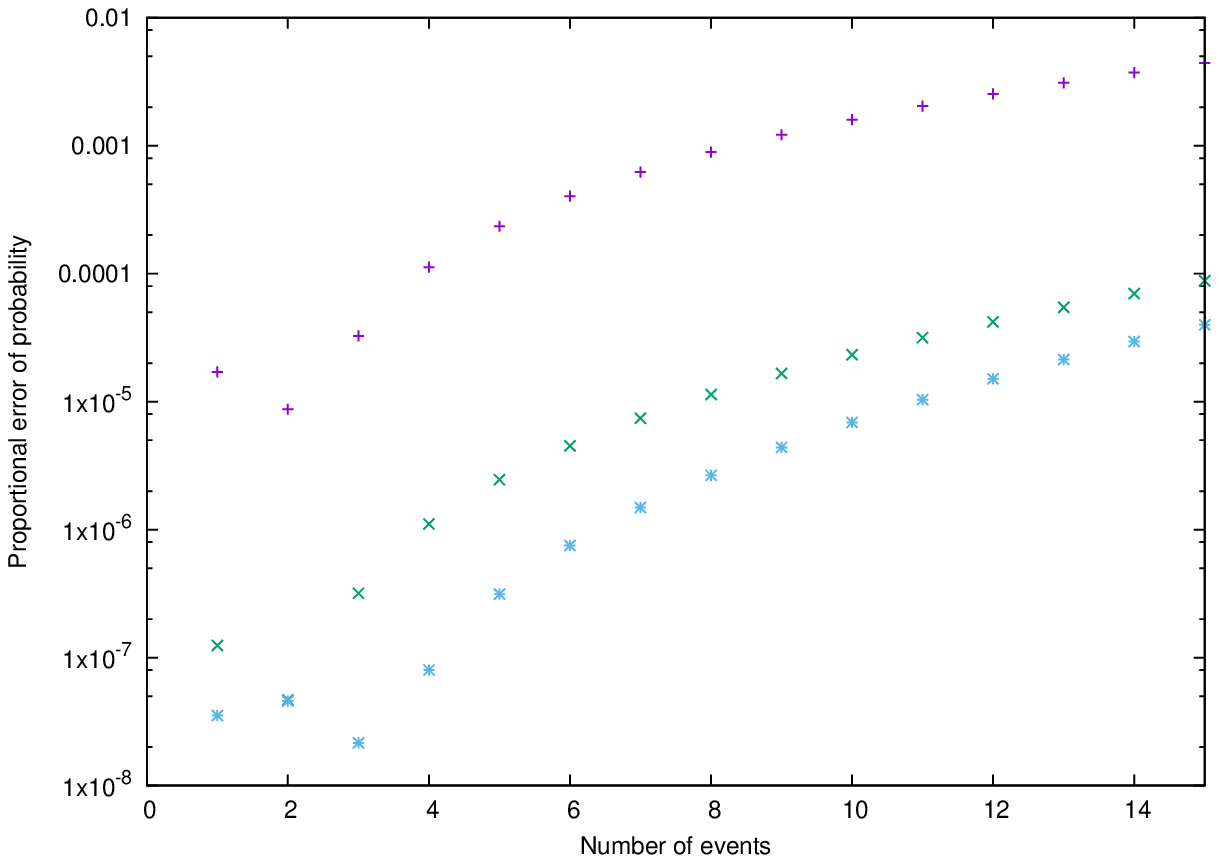}}
\caption{\label{figc} Proportional errors in probabilities for the \naive computation and the two Richardson corrections.
Here $\alpha=1, t=1, \beta=0.3$.}
\end{figure}

Figure~\ref{figb} shows the 3 errors for $\beta=0.6$.
Here again the extrapolations progressively
reduce error. Figure~\ref{fige} shows the estimated powers of $h$
for the errors, where now the curves get higher after each extrapolation.
Here the initial power is about 1.6, because $1+\beta < 2$. It then increases to 2,
and after applying the second extrapolation, to around 2.6.
Finally, Figure~\ref{figc} shows that the extrapolation works even for a low $\beta=0.3$.

\section{Generalisations}
\label{sec:general}
The methodology applies with no change (except the function that provides the
survival function) to some generalisations of the Weibull distribution.
Thus making the scale $\alpha^\beta$ a gamma random variate leads to the
distribution (See \citep[Section 3.1, page 374]{shane} for more details on
the derivation) with survival function
\[
S(t) = \frac{1}{(1 + (\alpha t)^\beta)^{\nu}},
\] where $\nu > 0$. 
This is the Burr type \rom{12} distribution where $\alpha$ is the scale parameter
and $\beta$ and $\nu$ are the shape parameters. When $\beta=1$ reduces to the
Lomax distribution (a shifted Pareto distribution). When $\nu=1$ this is
the log-logistic distribution, and as $\nu\rightarrow\infty$
we regain the Weibull distribution. This distribution addresses the problem of
heterogeneity.

The algorithm described can also cope with many of the Weibull-based distributions
described in \citet{lai}. It also copes with the gamma distribution,
where a function for the gamma survival function is needed. Here of course,
an analytic solution is available. Another interesting distribution that could be
used with the convolution method is the generalised gamma first introduced by
\citet{stacy1962generalization}. This distribution includes the Weibull, gamma
and log-normal as special cases. \citet{prentice1974log} proposed an alternative
parametrisation which is preferred for computation. In the \citet{prentice1974log}
parametrisation, the distribution has three parameters $(\mu, \sigma, q)$, and
its survival function is given by:

\begin{equation*}
  S(t) =
  \begin{cases}
     1 - I(\gamma, u) \ & \   \text{if q} > 0 \nonumber \\ 
     1 - \Phi(z) \ & \ \text{if q} = 0      \nonumber
\end{cases}
\end{equation*}

where $ I(\gamma, u) = \int_0^u x ^{\gamma - 1} \exp(-x) / \Gamma (\gamma) $ is the
regularised incomplete gamma function (the  gamma distribution function with
shape $ \gamma $ and scale $1$), $ \Phi$ is the standard normal distribution function,
$ u = \gamma \exp(|q|z), z = (\log(t) - \mu) / \sigma $, and $\gamma = 1 / q^2$. 

More generally, the convolution step can be applied to any survival distribution.
The Richardson improvement of Section \ref{sec:Richar} requires one to study the
first step error to derive a relevant extrapolation. Nevertheless, this extrapolation
can be skipped if one is willing to opt for a 'finer' convolution (and hence
inevitably longer computation times).

As mentioned in the introduction, the method can also be applied to a modified or delayed renewal process,
where the time to the first event follows a different distribution,
with pdf $f_0(x)$. This is useful for modelling distributions where
the percentage of zero events is abnormal, and one uses zero-inflated and hurdle models.
When both distributions are exponential, we obtain the `burnt fingers' distribution
of Greenwood and Yule \citep{john}. For the general case, it is straightforward to tweak
the code for finding single probabilities. The algorithm is:
\begin{enumerate}
\item if $m$ is 0, return the survival function derived from $f_0$;
\item if $m$ is 1, convolve $f_0$ with $P_0$ using (\ref{eq:step2});
\item for higher $m$, find $f_{m-1}(u)$ using the previous code,
  then convolve this with $f_0$ and finally apply (\ref{eq:step2}).
\end{enumerate}
Note that the convolution method can be readily extended to allow modified renewal processes, whereas series-expansion methods cannot.
\section{Estimation and testing}
\label{sec:ex}
\subsection{Data}
\begin{figure}
  \centering
  \makebox{\includegraphics[width = 0.6\textwidth]{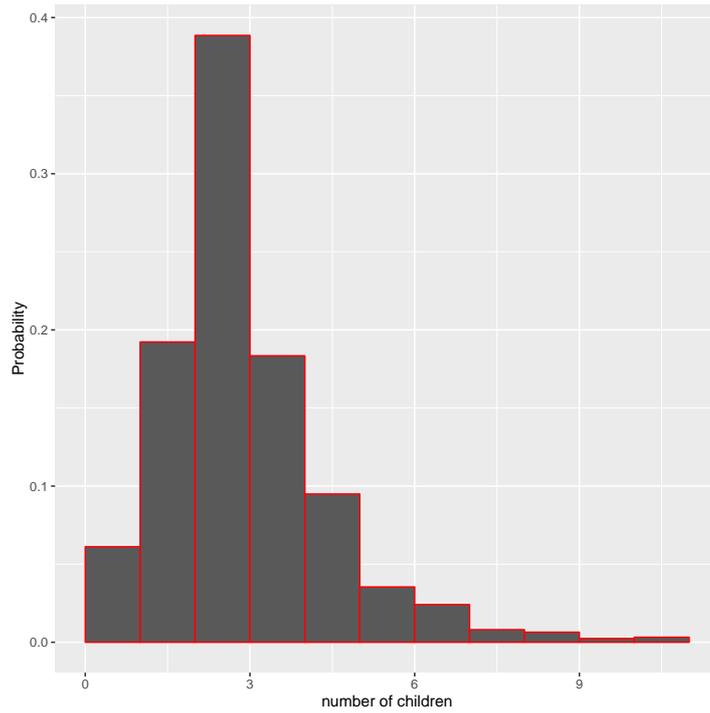}}
  \caption{\label{fig:histFertility}Frequency distributions of
    the number of children born to a woman
    who has completed childbearing in Germany ($n = 1,243$)}
\end{figure}

To illustrate the different algorithms described earlier as well as methods
previously suggested in the literature, we use a data set for completed
fertility. Completed fertility refers to the total number of children born
to a woman who has completed childbearing. The data set considered is
the same as the one analysed by \citet{winkl} and \citet{shane} and
consists of a sample of $n = 1,243$ women over 44 in 1985. A more detailed
description can be found in \citet{winkl}. We selected this
data set for two main reasons. First, the previous references inspired this
research and will be used as a benchmark for our new approach. It was essential to
be able to produce results in agreement with previous conclusions and hence
re-analysing the same data made sense. Second, this data set is slightly
underdispersed (sample variance $ 2.3$ versus the sample mean $2.4$) and hence allows us
to highlight the flexibility of the new approach compared to
the simple Poisson-negative binomial methods.
A more precise description of the data is presented in Figure~\ref{fig:histFertility}
and Table \ref{tab:freqTab1}.

The range of the data is quite narrow, with more than 95\% of the observations in the
range 0-5 and the highest count being 11 in both cases. The data set shows a
pronounced mode at 2 children, a number seen as ideal by many families.
\subsection{Comparing  performance of different methods}
In this section, we compare the  performance of the various methods using the German fertility
data and a univariate Weibull count model, intercept-only. We computed the model
log-likelihood by a very long convolution (20,000 steps as before), and proportional
errors computed taking this as correct after Richardson extrapolation. For each method,
we achieved the minimum number of computations to reach an precision (error) of at least
$10^{-8}$. The computation was repeated 1000 times and execution times measured using
routines from the \R package \texttt{rbenchmark}. The experience was
conducted on a 2.6 GHz intel Core i7 computer and results are collected in
Table~\ref{tab:perf1}.

\texttt{series-Euler-van} is the series expansion method accelerated
by the Euler and van-Wijngaarden transformations, \texttt{series-mat}
is the series expansion as described in \citet{shane} programmed in
vectorized form, \texttt{direct} is the direct convolution algorithm
described in Section~\ref{sec:allProba} and \texttt{naive} and \texttt{De Pril}
are described in Section~\ref{sec:proba}. Convolution methods are tested with and
without Richardson extrapolation. Table~\ref{tab:perf1} suggests that the series
expansion methods are almost twice as fast as the convolution methods and
more than 5 times faster than convolutions without Richardson correction.
Surprisingly, the De Pril method (with correction) performed slightly worse than
the direct approach and similarly to the naive approach. The reason is that this
method needed slightly more steps to reach the desired accuracy.
\footnote{When extrapolation was applied, the De Pril approach needed 36 steps when
  the other methods required only 24. If no extrapolation was applied, all methods
  used 132 steps. In this case, the De Pril method was found to be faster (32 \% faster
  than the naive approach and 53 \% faster compared to the direct approach.}
However, the De Pril method has been found to be slightly more accurate than all
other methods including series expansion for large counts (larger than 10). Given
that the testing data set we use has a narrow range of (low) counts, the
added value of the method was not seen.

In order to highlight the improvement introduced by the De Pril approach, we slightly
modified the German fertility data set by 'artificially' adding some large counts.
The new data is summarised in Table~\ref{tab:freqTabSim} and the new
performance in Table~\ref{tab:perfTabSim}. The results are more accordance
with what we expect. The De Pril approach is three times faster
than the naive approach and more than four times faster than the direct approach.
Nevertheless, it is still slower than the series approach
(the accelerated approach still being slightly faster than the vectorial approach).
It is not surprising that a 'tailored' method such as the series
expansion outperforms a generic method such the convolution method described in this
paper. Nevertheless, computation times are comparable and the convolution approach
has the advantage of being more much flexible as it allows any survival
distribution, and can be adapted for modified renewal processes. One pays the price for this flexibility in slightly increased computation time.

\subsection{Univariate models}
The first family of models considered is an intercept-only
(no individual covariates) version of several renewal processes 
with different distributions for the inter-arrival times.
Table ~\ref{tab:GermanUnivariate} presents values of model-choice criteria for the various models.

First, we note from Table~\ref{tab:freqTab1} that the Poisson model over-fits
the zero count and under-fits the peak at 2.

The log-likelihood values reported in Table~\ref{tab:GermanUnivariate} show best fit
by the generalised gamma, which is clearly preferred according to AIC and BIC.
Significant improvments are comfirmed by likelihood ratio tests over
Poisson (-2LR = 39.2) and gamma (-2LR = 30.7) at any convential level of
significance. The result is similar for the Weibull process model (-2LR = 26.3) compared
with Poisson. It is also worth mentioning here that the chi-squared goodness of fit test
rejects the null hypothesis (that the empirical data comes from the claimed distribution)
at any convential level of significance for the four models suggesting that
these simple models (with no covariates) fail to capture the data generating process.
A closer investigation of the table of observed and expected frequencies tells us
that all models under-estimate the pick at 2 childreen. Nevertheless, as mentioned earlier,
it made sense to analyse this dataset in order to be able to validate and compare
the results to what have been suggested in the literature.

One can also note that the log likelihood value presented in
Table~\ref{tab:GermanUnivariate} computed with the convolution method is identical to
the one in \citet[Table 1]{winkl} and \citet[Table 1]{shane}, thus validating
the accuracy of our computation. The standard errors are obtained
from numerical computation of the Hessian matrix at the fitted value of the
parameters.

\subsection{Regression models using renewal processes}
We turn now to the analysis of the model with individual covariates.
The explanatory variables available are the woman's general education
(given by the number of years of school), nationality (a dummy, either German
or not), university access (yes or no), rural or urban dwelling, religion
(a categorical variable with levels Catholic, Protestant, and Muslim, with
others being the reference group), year of birth and the year of marriage).
Results are collected in Table~\ref{tab:GermanReg}. 

One can also note here that the values of the log likelihood are in accordance
with the previously mentioned literature. The value of the coefficients are not
exactly identical but are within the same confidence region. The
generalised gamma distribution still provides the best likelihood, but with a higher AIC, so
the Weibull model would be (slightly) preferred. One may conclude that
the introduction of individual covariates improves the data description rather more
than a more flexible hazard model (as introduced by the generalised gamma). 

We would also like to mention here that we tried to reproduce the
heterogeneous-gamma described in \citet[Table 2]{shane}. We found similar results
using the series expansion methods when we used 50 terms to expand the series but
different results were obtained (with smaller log-likelihood values) when more terms
were used. We think that the series expansion may need more then 50 terms to converge
in the heterogeneous-gamma case and hence the conclusion of \citet[Table 3]{shane} should be
interpreted with care. Although the series expansion method works smoothly in the simple
Weibull case (around 20 terms are usually enough to ensure convergence), for more
complicated distribution such as the Weibull-gamma more terms may be needed.
On the other hand, due to the use of the gamma function, there is a limitation
on the maximum number of terms that could be numerically computed. The convolution
method described in this paper does not suffer from this limitation and hence can
be seen as more robust as well as being more flexible. 

\section{Conclusions}
\label{sec:cl}
A fast and flexible method is presented for computing the probabilities of discrete distributions derived from renewal and modified renewal processes.
This should pave the way for more widespread use of this type of model in econometrics, health science, and wherever count data needs to be modelled.
Where the data arise from a stochastic process, such as football goals or hospital visits, the renewal model can have a strong basis in fact.
It can however be applied  to any count data, such as number of bacteria seen under a microscope, using the renewal framework purely as a mathematical device.

This class of models is we think tractable enough for use by practitioners. Computation of probabilities of numbers of events is essential for likelihood-based inference,
and we have focused on this. Tests are often also needed, \eg for under or overdispersion. If fitting a Weibull model, as the shape parameter $\beta$ determines under or overdispersion,
we simply need to test that $\beta=1$. Computing the log-likelihood with $\beta$ `floating' and fixed to unity, twice the increase in log-likelihood on floating $\beta$
is asymptotically distributed as $X^2[1]$, a chi-squared with one degree of freedom. For small samples, one can find the distribution of this statistic
under $H_0$ more accurately by using the parametric bootstrap. We would thus claim that these distributions are tractable where it matters: computation of moments for example is difficult, but is not needed for inference.
We would suggest that a Monte-Carlo simulation would be easy to program and fast enough for the modest accuracy required.

We have chosen to implement what seemed the most direct method of computing probabilities, after ruling out Monte-Carlo integration on the grounds that regular quadrature methods are better
for one-dimensional integrals. The method given can be applied as it stands to a variety of generalisations of the Weibull distribution, and can be applied in outline to other survival distributions, such as the lognormal.
An R package that allows the Weibull, gamma and few other distributions is available
from the \texttt{CRAN} archive.

This is an area where much further work could be done. There is a bewildering variety of possible approaches to computing the probabilities,
and the successful use of Laplace or Fourier transforms is surely a possibility. However, the disadvantage of direct methods, that computation time goes as $N^2$ for $N$ steps,
is much ameliorated by using Richardson extrapolation, so that $N$ can be small. 
The Weibull distribution has a virtue for the direct convolution approach adopted here, in that the distribution function is easy to compute.
However, it has the disadvantage for transform methods that the transform $M(s)$ cannot be found analytically, but must be evaluated numerically for each value of $s$,
where the transform is $M(s)=\int_0^\infty \exp(-st)\dee F(t)$. The present method, which already gives adequate performance,  would be a useful benchmark for developers of more advanced methods to compare with.
We conjecture that great improvements in speed are not possible, but hope to be proved wrong here.

Perhaps of greater interest than further speeding up computation is gaining experience with the expanded range of renewal-type models that can now be feasibly used.
This includes modified renewal processes, where the time to the first 
event follows a different distribution to later events. This for example yields a natural class of hurdle models, where the first event is slow to happen, but later events follow more quickly.
Conversely, this class includes distributions where there are very few occurrences of zero events. It will be interesting to see how useful practitioners find these new models.

\bibliographystyle{apalike}
\bibliography{REF}

\newpage
\begin{appendices}

\section{Addition chain method for computing probabilities}
\label{appendix:ch2A}
The aim is to find the $m$th convolution of the pdf in as few convolutions as possible.
The method works by convolving the pdf $f_i$ of 
$i$ events occurring, using 
\begin{equation}
f_{i+j}(t)=\int_0^t f_i(u)f_j(t-u)\dee u\
\label{eq:step1}
\end{equation}
and finally
\begin{equation}
P_m(t)=\int_0^tf_m(u)P_0(t-u)\dee u
\label{eq:step2}
\end{equation}
We need two work arrays: one (pdfn) for the $n$-th convolution of the pdf, initially set 
to $\text{pdfn}[j]=(F((j-1)h)-F(jh))/h$, an approximation to $f_1$,
and repeatedly overwritten, the other, q, to hold what will become the final pdf as it 
is being updated. Two routines are needed to do the convolving: one for convolving the 
$m$th order pdf with itself, the other for convolving two pdfs of different order.
The symmetry of the integrand means that only half the multiplications are required when 
doubling the order of the pdf.

To organize the calculation, we first find the binary decomposition of $m$. For example, 
with $m=21$, we would have $21=1+2^2+2^4$. This can be translated into code as: 
\begin{itemize}
\item set $q$ to $f_1$, 
\item apply (\ref{eq:step1}) to obtain $f_2$,
\item then apply (\ref{eq:step1}) to $f_2$ to obtain $f_4$, 
\item convolve $q$ with $f_4$ to obtain q as $f_5$,
\item apply (\ref{eq:step1})again to $f_4$ to obtain $f_8$ and $f_{16}$,
\item then convolve q with $f_{16}$ to obtain $f_{21}$.
\item Finally, apply (\ref{eq:step2}) to obtain $P_{21}(t)$.
\end{itemize}
This has required 6 convolutions and one evaluation, instead of 20 convolutions 
and one evaluation.

The best case occurs when $m=2^k$, when $k$ convolutions are needed, all order doublings. 
The worst case occurs when $m=2^k-1$, when $m=\sum_{j=0}^{k-1} 2^j$.
Here all the pdfs $f_1, f_2 \cdots f_{k-1}$ must be convolved, 
giving a total of $2(k-1)$ convolutions. This is still $O(\ln_2(m))$.
 
\section{Richardson extrapolation}
\label{appendix:ch2B}
This technique can substantially reduce the required number of steps $N$. To derive a useful extrapolation
we start by considering the error of the extended midpoint approximation.
The error $E_j$ is given by
\begin{equation}
E_j = \int_{(j-1)h}^{jh}g(u)\dee F(u) - g\{(j-1/2)h\}(F\{jh\}-F\{(j-1)h\}).
\label{eq:ej}
\end{equation}
Expanding the integrand in a Taylor series 
$g(u)\simeq g(u_0) + g^{\prime} \ldotp (u-u_0) +
(1/2) g^{\prime \prime} \ldotp (u-u_0)^2$, 
where $u_0=(j-1/2)h$ and the derivatives are taken at $u_0$.
Writing similarly the pdf 
$f(u) = f(u_0) + f^{\prime} \ldotp (u-u_0) + (1/2) f^{\prime \prime} \ldotp (u-u_0)^2$, 
we have for the step error to the lowest order in $h$,
\begin{equation}
E_j=h^3\{f^{\prime}g'/12 + fg^{\prime \prime}/24 \}.
\label{eq:ejFom}
\end{equation}
The proof follows:

We have that
\[ g(u)  \simeq  g(u_0) + g^{\prime} \ldotp (u-u_0) + (1/2) g^{\prime \prime} \ldotp (u-u_0)^2,\]
so that 

\begin{subequations}
\begin{align}
  E_j & = & \int_{(j-1)h}^{jh}g(u)\dee F(u) - g\{(j-1/2)h\}(F\{jh\}-F\{(j-1)h\}) \nonumber \\
      & = & \int_{(j-1)h}^{jh}(g^{\prime} \ldotp (u-u_0) + (1/2) g^{\prime \prime} \ldotp (u-u_0)^2)f(u)\dee u. 
  \label{eq:interm}
\end{align}
\end{subequations}

Expanding 
\[ f(u)  \simeq  f(u_0) + f^{\prime} \ldotp (u-u_0) + (1/2) f^{\prime \prime} \ldotp (u-u_0)^2\]
and substituting in (\ref{eq:interm}) we obtain
\[E_j \simeq \int_{(j-1)h}^{jh}\{g^{\prime} \ldotp (u-u_0) + (1/2) g^{\prime \prime} \ldotp (u-u_0)^2\}\{f(u_0) + f^{\prime} \ldotp (u-u_0) + (1/2) f^{\prime \prime} \ldotp (u-u_0)^2\}\dee u.\]
The integrand $I(u)$ is:

\begin{subequations}
\begin{align}
I(u) & \simeq & g(u_0)  & (u - u0) \{ g^\prime \ldotp f(u_0)\} \label{eq:term1}\\ 
                   & + & (u - u0)^2 \{ g^\prime \ldotp f^\prime + 
                       (1/2) g^{\prime \prime} \ldotp f(u_0) \} \label{eq:term2}
\end{align}
\end{subequations}
Then we need to integrate each term in the previous equation between $(j-1)h$ and $jh$:
\begin{itemize}
\item Integration of Equation~(\ref{eq:term1}) gives 0 by symmetry.
\item Integration of Equation~(\ref{eq:term2}) gives $h^3/12 \times
\{ g^\prime \ldotp f^\prime + (1/2) g^{\prime \prime} \ldotp f(u_0)\}$
\end{itemize}
Therefore, using the definition of $Ej$ in (\ref{eq:ej}), we get the result in Equation~(\ref{eq:ejFom}).

Since there are $N=t/h$ terms, this gives an error of $O(h^2)$. 
However, the first step cannot be treated in this way, because $u^\beta$ 
has a singularity at $u=0$, which is therefore at the radius of convergence 
of the Taylor expansion. 
We instead consider the the error of the first term when $F(u)$ is 
approximated as $(\alpha u)^\beta$, \ie at small times $u$. Then the error $E_1$ can
be found from (\ref{eq:ej}) without expanding out $f$ as
\[
E_1   \simeq
      g^{\prime} k_1(\beta) (\alpha h)^{\beta + 1} / \alpha  +  
      g^{\prime \prime} k_2(\beta) (\alpha h)^{\beta + 2} / \alpha^2 
\]
where $k_1(\beta)$ and $k_1(\beta)$ are some functions of $\beta$ that could be found 
exactly. This is $O(h^{\beta + 1})$. For $\beta > 1$, the $O(h^2)$ error dominates, 
but for $\beta < 1$ the error is $O(h^{\beta+1})$. Higher order errors are of 
type $O(h^{\beta+n})$ and $O(h^{n\beta+1})$ for $n > 1$.

This affects what can be achieved by Richardson extrapolation. 
Two steps are advocated using (\ref{eq:rich}), so that 3 sets of convolutions are done
with series lengths $N, 2N, 4N$. Let a particular probability be $A_1, A_2$ and $A_3$ 
from the convolutions (in order of increasing length).
Then the extrapolation used is: Define $\gamma_1 = \beta+1$, $\gamma_2=2$ 
(the order does not matter). 
Compute $B_1=(2^{\gamma_1} A_2-A_1)/(2^{\gamma_1}-1)$,
$B_2=(2^{\gamma_1} A_3-A_2)/(2^{\gamma_1}-1)$.
Finally, the extrapolated probability is $C_1=(2^{\gamma_2} B_2-B_1)/(2^{\gamma_2}-1)$. 
We have removed the two errors, leaving higher order errors:
$O(h^{\beta+2})$ and $O(h^4)$.
When $\beta > 1/2$, two extrapolations leave an error of order 
$\text{min}(1+2\beta, 2+\beta, 4)$, which is at least $O(h^3)$. 
When $\beta$ is small, say 0.1, there are many errors of similar orders, 
and Richardson extrapolation, although it can improve accuracy, can not remove 
the low-order error. However, we believe that the procedure recommended will generally be
satisfactory, and if not, for low $\beta$ one would have to increase $N$.

The code that carries out the extrapolation also computes the minimum number 
of exponentiations, because some of those for $4N$ can be re-used for $2N$ and $N$.

For studying the order of error, a very long convolution was used, with 20000 steps,
and errors computed taking this as correct (after Richardson extrapolation).
The order of error can be studied by carrying out three convolutions with 
$N, 2N, 4N$, and solving the 3 equations for $\gamma$. We then find
\begin{equation}\gamma=\ln\frac{S_2-S_1}{S_3-S_2}/\ln(2),\label{eq:gamma}\end{equation}
where $S_1=S+ah^\gamma$ etc. The extrapolated value $S_1^e$ is 
\[S_1^e=\frac{S_1S_3-S_2^2}{S_1+S_3-2S_2}.\]
This is in fact the `Aitken acceleration' of $S_1$, sometimes used to speed up
convergence of series, where $S_1, S_2, S_3$ would be successive partial sums.
\citet{press} recommend writing it in the form
\begin{equation}S_1^e=S_1-(S_1-S_2)^2/(S_1+S_3-2S_2),\label{eq:s}\end{equation} 
which reduces rounding error.

Although this extrapolation improves the results when
$\beta < 1$, the procedure recommended is sometimes more accurate.
However, one could use either. It can be seen from (\ref{eq:s})
that unlike the recommended procedure, longer convolutions
do not have more weight, and that there is the potential
for divide overflow and loss of accuracy in computing $S$.

It is possible in the same way to go further, and remove the next power of error,
$\beta+2$. Equation (\ref{eq:gamma}) was applied to the probabilities $C_1, C_2, C_3$.
This requires 5 initial computations, of $A_1\cdots A_5$. The power of $h$ remaining was
roughly $\beta+2$, but decreased below this when $\beta < 0.5$.
However, application of the Richardson extrapolation will reduce error,
even if the power of $h$ used, $\gamma_2$, is not correct,
and the true power is $\gamma_1$. It is easy to show that error is reduced
if $\gamma_2 \ge \gamma_1$. Hence this third Richardson step  will always
reduce the error further.
\end{appendices}

\newpage
\section*{Tables and Figures}
\begin{figure}
\centering
\makebox{\includegraphics{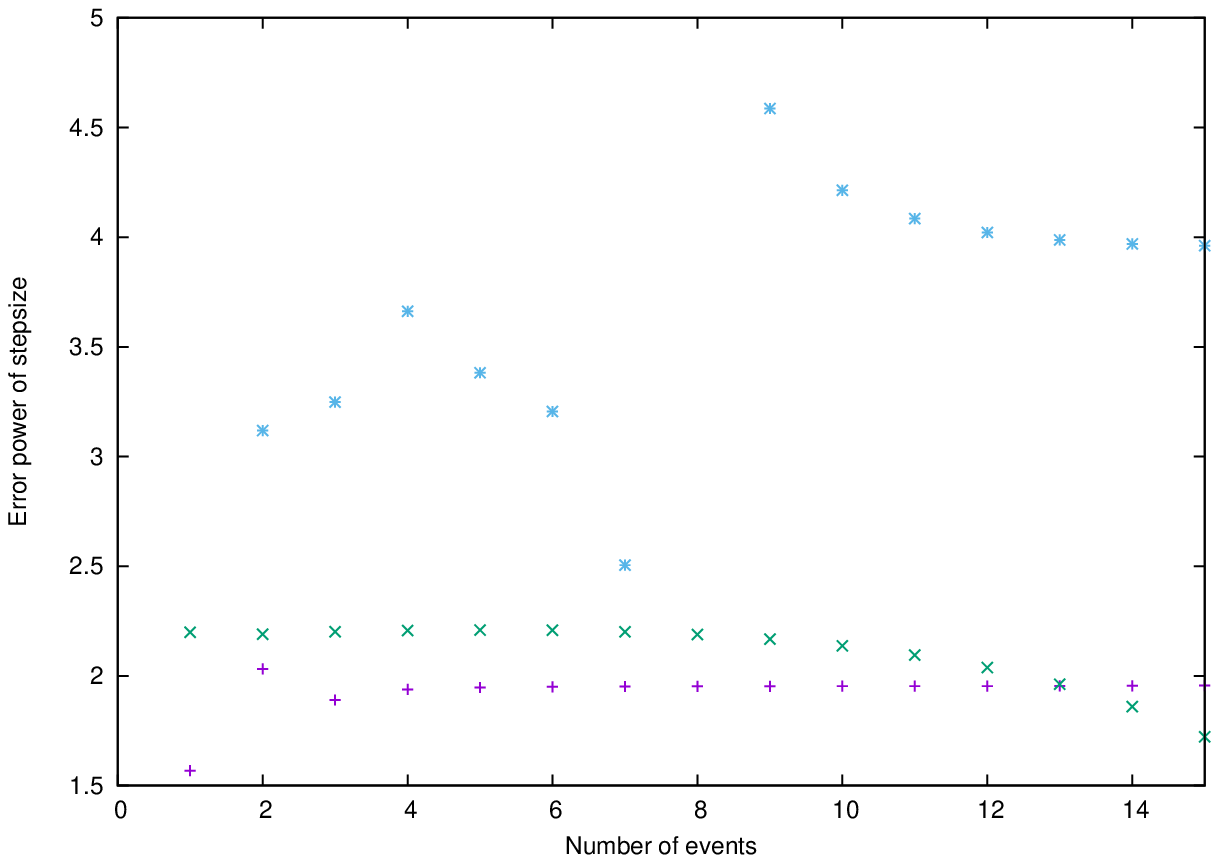}}
\caption{\label{figd} Powers of stepsize $h$ for error  in probabilities for the \naive computation and the two Richardson corrections.
Here $\alpha=1, t=1, \beta=1.2$.}
\end{figure}

\begin{table}[!htb]
    \centering
    \begin{tabular}{ccccccccccccc}
      \toprule
      Children & 0 & 1 & 2 & 3 & 4 & 5 & 6 & 7 & 8 & 9 & 10 & 11 \\ 
      \midrule
      Frequency & 76 & 239 & 483 & 228 & 118 & 44 & 30 & 10 & 8 & 3 & 3 & 1 \\ 
      Percent & 6.1 & 19.2 & 38.9 & 18.3 & 9.5 & 3.5 & 2.4 & 0.8 & 0.6 & 0.2 & 0.2 & 0.1 \\
      Poisson fitted & 9.2 & 21.9 & 26.2 & 20.8 & 12.4 & 5.9 & 2.3 & 0.8 & 0.2 & 0.1 & 0.0 & 0.0 \\ 
      \bottomrule
    \end{tabular}
    \caption{\label{tab:freqTab1}Number of children in the German fertility dataset.}
\end{table}

\begin{table}[!htb]
    \centering
    \begin{tabular}{c|cc}
     \toprule
      method & relative & elapsed (in seconds) \\ 
      \midrule
      series-Euler-van  & 1.00 & 19.86 \\ 
      series-mat  & 1.09 & 21.74 \\ 
      direct-conv-extrapolation  & 1.82 & 36.09 \\ 
      naive-conv-extrapolation  & 1.93 & 38.29 \\ 
      De Pril-conv-extrapolation  & 1.93 & 38.40 \\ 
      De Pril-conv  & 5.73 & 113.72 \\ 
      naive-conv  & 7.57 & 150.30 \\ 
      direct-conv  & 8.76 & 173.98 \\ 
      \bottomrule
    \end{tabular}
    \caption{\label{tab:perf1}
    Performance measure of the different computation methods available for the
  Weibull count (German fertility data). The methods are described in the main text.}
\end{table}

\begin{table}[ht]
\centering
\begin{tabular}{c|ccccccccccccccc} 
  \toprule
  Children & 0 & 1 & 2 & 3 & 4 & 5 & 6 & 7 & 8 & 9 & 10 & 11 & 12 & 13 & 14 \\
  \midrule
  Frequency & 76 & 239 & 483 & 228 & 118 & 44 & 30 & 10 & 53 & 43 & 59 & 44 & 50 & 45 & 56 \\ 
  Percent & 4.8 & 15.1 & 30.6 & 14.4 & 7.5 & 2.8 & 1.9 & 0.6 & 3.4 & 2.7 & 3.7 & 2.8 & 3.2 & 2.9 & 3.5 \\
  \bottomrule
\end{tabular}
\caption{
    \label{tab:freqTabSim}
    Number of children (simulated data with artificially larger count)
  }
\end{table}

\begin{table}[!htb]
    \centering
\begin{tabular}{c|cc}
  \toprule
 method & relative & elapsed (in seconds) \\ 
  \midrule
  series-Euler-van & 1.00 & 36.69 \\ 
   series-mat & 1.03 & 37.77 \\ 
  De Pril-conv-extrapolation & 3.49 & 128.11 \\ 
  naive-conv-extrapolation & 10.13 & 371.68 \\ 
  De Pril-conv  & 10.94 & 401.48 \\ 
  direct-conv-extrapolation  & 13.64 & 500.33 \\ 
  naive-conv  & 113.13 & 4150.35 \\ 
  direct-conv  & 233.15 & 8553.23 \\ 
   \bottomrule
\end{tabular}
\caption{\label{tab:perfTabSim}
    Performance measure of the different computation methods available for the
    Weibull count model (simulated data set)}
\end{table}

\begin{table}[ht]
\centering
\begin{tabular}{ccccccccc}
  \toprule
    & \multicolumn{2}{c}{Poisson} & \multicolumn{2}{c}{Weibull} &
    \multicolumn{2}{c}{Gamma} & \multicolumn{2}{c}{gen. Gamma} \\
    \cmidrule(lr){2-3}
    \cmidrule(lr){4-5}
    \cmidrule(lr){6-7}
    \cmidrule(lr){8-9}
    Variable & Coef & SE & Coef & SE & Coef & SE & Coef & SE \\
    \midrule
scale & 2.38 & 0.02 & 2.64 & 0.03 & 0.35 & 0.06 & 0.64 & 0.09 \\ 
  shape &  &  & 1.12 & 0.03 & 1.16 & 0.06 & 1.93 & 0.07 \\ 
  shape2 &  &  &  &  &  &  & 2.29 & 0.38 \\
  \midrule
  log likelihood & -2186.78 &  & -2180.36 &  & -2182.53 &  & -2167.18 &  \\ 
  AIC & 4375.55 &  & 4364.71 &  & 4369.06 &  & 4340.37 &  \\ 
  BIC & 4380.68 &  & 4374.97 &  & 4379.31 &  & 4355.74 &  \\
  \midrule
  $\chi^2$ & 126.16 & & 111.79 & & 115.53 & & 87.29 & \\
  df  & 6  & & 5 & & 5 & & 4 &  \\
  p-value & $8.2 \times 10^{-25}$ & & $1.7 \times 10^{-22}$& & $2.7 \times 10^{-23}$& & $4.9 \times 10^{-18}$& \\
  \bottomrule 
\end{tabular}
\caption{\label{tab:GermanUnivariate}
    German fertility data: Model choice criteria for the various models.
  }
\end{table}

\begin{table}[ht]
  \centering
  \begin{tabular}{ccccccccc}
    \toprule
    & \multicolumn{2}{c}{Poisson} & \multicolumn{2}{c}{Weibull} &
    \multicolumn{2}{c}{Gamma} & \multicolumn{2}{c}{gen. Gamma} \\
    \cmidrule(lr){2-3}
    \cmidrule(lr){4-5}
    \cmidrule(lr){6-7}
    \cmidrule(lr){8-9}
    Variable & Coef & SE & Coef & SE & Coef & SE & Coef & SE \\
    \midrule
    scale & 3.150 & 0.302 & 4.044 & 0.315 & 0.211 & 0.252 & -1.087 & 0.252 \\ 
    German & -0.200 & 0.072 & -0.223 & 0.072 & -0.190 & 0.059 & -0.190 & 0.059 \\ 
    Years of schooling & 0.034 & 0.032 & 0.039 & 0.033 & 0.032 & 0.027 & 0.032 & 0.026 \\ 
    Vocational training & -0.153 & 0.044 & -0.173 & 0.044 & -0.144 & 0.036 & -0.144 & 0.036 \\ 
    University & -0.155 & 0.159 & -0.181 & 0.160 & -0.146 & 0.130 & -0.146 & 0.129 \\ 
    Catholic & 0.218 & 0.071 & 0.242 & 0.070 & 0.206 & 0.058 & 0.206 & 0.058 \\ 
    Protestant & 0.113 & 0.076 & 0.123 & 0.076 & 0.107 & 0.062 & 0.107 & 0.062 \\ 
    Muslim & 0.548 & 0.085 & 0.639 & 0.087 & 0.523 & 0.070 & 0.523 & 0.069 \\ 
    Rural & 0.059 & 0.038 & 0.068 & 0.038 & 0.055 & 0.031 & 0.055 & 0.031 \\ 
    Year of birth & 0.002 & 0.002 & 0.002 & 0.002 & 0.002 & 0.002 & 0.002 & 0.002 \\ 
    Age at marriage & -0.030 & 0.007 & -0.034 & 0.006 & -0.029 & 0.005 & -0.029 & 0.005 \\ 
    shape &  &  & 1.236 & 0.034 & 1.439 & 0.071 & 2.211 & 0.031 \\ 
    shape2 &  &  &  &  &  &  & 1.121 & 0.169 \\
    \midrule
    log likelihood & -2101.8 &  & -2077.0 &  & -2078.2 &  & -2076.7 &  \\ 
    AIC & 4225.6 &  & 4178.0 &  & 4180.5 &  & 4179.6 &  \\ 
    BIC & 4281.980 &  & 4240 &  & 4242 &  & 4246.2 &  \\ 
    \bottomrule
  \end{tabular}
  \caption{\label{tab:GermanReg}
    Regression model results for German fertility data
  }
\end{table}

\end{document}